%% file: sv.tex
\let\origvec\vec
\documentclass{llncs}
\let\springervec\vec
\let\vec\origvec

\usepackage[english]{babel}
\usepackage[utf8]{inputenc}
\usepackage[T1]{fontenc}
\usepackage{graphicx}
\usepackage{amsmath}
\usepackage{amsfonts}
\usepackage{amssymb}
\usepackage{graphicx}
\usepackage[caption=false]{subfig}
\usepackage[plainpages=false]{hyperref}
\usepackage{enumitem}

\let\origvec\vec
\let\vec\springervec

\usepackage{textcomp}

\newcommand{\card}[1]{\ensuremath{\left|#1\right|}}
\newcommand{\myspace}[1]{\begin{minipage}[b][#1]{\textwidth}\end{minipage}}
\begin{document}
\title{Crossing Minimization in Storyline Visualization}

\author{
Martin Gronemann\inst{1}, 
Michael J{\"u}nger\inst{1}, 
Frauke Liers\inst{2}, 
Francesco Mambelli\inst{1}
}
\authorrunning{M.~Gronemann, M.~J{\"u}nger, F.~Liers, F.~Mambelli}
\tocauthor{M.~Gronemann, M.~J{\"u}nger, F.~Liers, F.~Mambelli}

\institute{
Department of Computer Science, University of Cologne\\
\email{\{gronemann,mjuenger,mambelli\}@informatik.uni-koeln.de}
\and
Department of Mathematics, University of Erlangen-N{\"u}rnberg\\
\email{frauke.liers@math.uni-erlangen.de}
}

\maketitle

\pagestyle{plain}
\input{abstract.tex}
\input{introduction.tex}
\input{modelling.tex}
\input{ilpformulation.tex}
\input{implementation.tex}
\input{computationalresults.tex}
\input{conclusion.tex}

\input{acknowledments.tex}

\bibliographystyle{splncs03}
\bibliography{references}

\end{document}

%% file: abstract.tex
\begin{abstract}
A storyline visualization is a layout that represents the temporal dynamics of social 
interactions along time by the convergence of chronological lines. Among the criteria 
oriented at improving aesthetics and legibility of a representation of this type, a small 
number of line crossings is the hardest to achieve. We model the crossing 
minimization in the storyline visualization problem as a multi-layer crossing minimization problem
with tree constraints. Our algorithm can compute a layout with the 
minimum number of crossings of the chronological lines. Computational results demonstrate 
that it can solve instances with more than $100$ interactions and with more than 
$100$ chronological lines to optimality.
\end{abstract}%

%% file: introduction.tex
\section{Introduction}\label{sec:Introduction}
Visualizing time-varying relationships between entities using converging 
and diverging curves on a timeline has received a considerable amount 
of interest recently. The ability to display interactions 
among entities, while at the same time being able to put these in a chronological 
context has found applications beyond its initial purpose which coined its name.
Munroe~\cite{xkcd} introduced the \emph{storyline visualization}
as hand-drawn illustrations in xkcd’s ``Movie Narrative Charts'', 
where lines represent the characters of various popular movies
and the scenes are ordered chronologically and represented 
by bundling the lines of the corresponding characters. This
concept has been used to visualize various spatiotemporal data,
like communities in time-varying graphs~\cite{egolines,communitylines}, 
software projects~\cite{softwarelines}, topic analysis~\cite{textflow}, etc.

However, hand crafted or semi-automated methods are limited in their applicability 
in a world of ever growing datasets. In order to obtain a storyline visualization 
automatically, Tanahashi and Ma~\cite{tm12} discuss various aspects of a 
well-designed storyline visualization and present an evolutionary algorithm 
that incorporates these in its objective function. They identify three important 
criteria that one usually wants to optimize: line crossings, whose number should 
be small, line wiggles, that should be avoided by drawing every chronological line 
as straight as possible, and space efficiency. Based on these aspects, 
Liu et al.~\cite{storyflow13} describe a technique which further improves 
the layout and runs significantly faster compared to the evolutionary algorithm in~\cite{tm12}.
Being able to create storyline visualizations of bigger instances, Tanahashi et al.~\cite{thm15} 
take this one step further and show how to create storyline visualizations from streaming data.

In this paper, we study the crossing minimization problem in storyline visualization 
from a combinatorial optimization point of view. While most approaches tackle this problem 
with heuristics, Kostitsyna et al.~\cite{knpss15} recently shed some light onto 
its combinatorial properties. Besides noting that the decision problem is NP-complete 
(by reduction from bipartite crossing number), they provide a lower bound for the number 
of crossings in a restricted variant of the problem and show that the general problem 
is fixed-parameter tractable in the number of characters. But a straightforward implementation 
of the algorithm is impractical, even for a small number of characters.

However, the problem is also similar to a few already 
well-studied problems in graph-drawing. It may seem 
that the problem is related to a special case of metro-line crossing minimization, 
in particular, the so called two-sided models in which metro-lines run only from left to right~\cite{bkps07,fp13}. 
However, all metro-line crossing minimization problems have in common that they are defined on a rail network 
whose embedding is fixed due to its geographical context. 
This difference makes a straightforward transformation difficult.

As already observed by Kostitsyna et al.~\cite{knpss15}, storyline crossing minimization has a strong
relationship to \emph{multi-layer crossing minimization (MLCM)}. Here each node of the graph is assigned to one 
of the layers (parallel straight lines) in such a way that each edge connects two vertices on consecutive layers. 
The aim is to find an ordering of the nodes on every layer such that the total number of edge crossings is minimized.
Although the corresponding planarity testing problem is linear-time solvable~\cite{jlm98}, 
the MLCM problem itself remains NP-hard even when restricted to two layers~\cite{gj83}. 
This led to the development of various heuristics, but also to exact approaches~\cite{bwz10,chjm11,hk04,jlmo97}.

In order to exploit existing techniques for solving MLCM instances, 
a straightforward transformation can be sketched as follows. We represent the characters
as paths in an MLCM instance, in which the layers mark important points in time, e.g.,~a new bundle has
to be created. Of course, a bundle of lines (paths) requires the corresponding vertices to be consecutive
on the layer, a constraint which is problematic in the general MLCM setting. 

But we can borrow ideas from another crossing minimization problem type, the so called \emph{tanglegrams}.
The general tanglegram problem consists of two trees and a set of edges
connecting the leaves of one tree with the leaves of the other, i.e., the leaves and the connecting edges form a bipartite graph. 
The objective is essentially to perform a bipartite (or two-layer) crossing minimization with the additional constraint
that leaves of the same subtree appear consecutively on the layers. 
However, when consulting the literature on tanglegrams,
attention must be paid to the details. Some definitions require the trees to be binary, while others
restrict the edge set to be a perfect matching, or both~\cite{Buchin2012,fernau2010,sascha09}.
Since the focus of the paper is not on tanglegrams, we restrict ourselves to the general case.
Here two works are of interest, Baumann et al.~\cite{bbl10} describe an ILP-based approach,
whereas Wotzlaw et al.~\cite{wsp12} employ a SAT-formulation.
However, not only the techniques differ, in~\cite{bbl10} only two layers are considered,
whereas the SAT approach in~\cite{wsp12} works on multiple layers but requires that the tree constraints are $k$-ary with $k > 1$ fixed.

The related problem of testing level planarity under tree constraints is discussed by Angelini~et~al.~\cite{addfr15}. They
show that if edges are restricted to run between consecutive layers, then the problem can be solved in quadratic time,
whereas if this restriction does not hold, the problem is NP-complete.

In this paper we solely focus on the crossing minimization 
problem in storyline visualization. Therefore, we neglect other design aspects
and restrict ourselves to the combinatorial problem, i.e., determining an 
ordering of the lines such that the number of crossings is minimum.
We model this problem as a special variant of the MLCM problem under tree constraints
and provide an ILP formulation for it. 
Computational results show that we are able to solve instances of moderate size to optimality 
within a few seconds. Moreover, we provide solutions for storyline instances from the literature,
some of which have been solved to optimality for the first time. 
These are of particular value, since they offer a reference when 
comparing the crossing minimization performance of heuristics.




%% file: modelling.tex
\section{Modelling Storyline Visualization as Multi-Layer Crossing Minimization with Tree Constraints}\label{sec:SVtoMLCM}

We begin with a formal definition of the multi-layer crossing minimization problem 
with tree constraints (MLCM-TC). The input for MLCM-TC consists of a graph $G=(V,E,\mathcal{T})$, 
where the set of the nodes $V=\bigcup_{r=1}^p V_r$ is partitioned into $p$ different layers. 
$E=\bigcup_{r=1}^{p-1} E_r$ is the set of the edges such 
that $E_r\subseteq V_r\times V_{r+1}$ for every $r\in\{1,2,\ldots,p-1\}$, 
i.e., each edge of $E_r$ has one end in $V_r$ and the other in $V_{r+1}$. 
$\mathcal{T}=\{T_{r}\mid r=1,2,\ldots,p\}$ is a family of rooted trees with at 
least one internal node (root node), whose leaves are exactly the nodes of $V_r$. 
In the following, whenever we consider a graph, we implicitly assume that it is of this type, 
which is known in the literature as ``(proper) $\mathcal{T}$-level graph''~\cite{addfr15}. 

Given an instance $G=(V,E,\mathcal{T})$ of MLCM-TC, the task is to determine, 
for each layer $r\in\{1,2,\ldots,p\}$, permutations $\pi_r=\langle v_1,v_2,\ldots,v_{\card{V_r}}\rangle$ 
of the nodes in $V_r$ such that for each internal node $\tau$ of $T_r$, 
all $t$ leaves in the subtree rooted at $\tau$ are adjacent in $\pi_r$, 
i.e., they form a sub-permutation $\langle v_i,v_{i+1},\ldots,v_{i+t-1}\rangle$ 
for some $i\in\{1,2,\ldots,\card{V_r}-t+1\}$.

An easy reduction of the NP-hard MLCM problem to the MLCM-TC problem 
(add a trivial tree with the root as the only internal node to each layer) shows that MLCM-TC is NP-hard. 
This justifies the usage of integer programming techniques in the next section. 
Now we give a formal description of the storyline visualization problem in order to 
support our hypothesis that \hbox{MLCM-TC} captures its core when the criteria ``line wiggle avoidance'' 
and ``space efficiency'' are neglected in favour of crossing minimization. 
A story consists of a set of characters $C=\{c_1,c_2,\ldots,c_n\}$ and 
a set of scenes $S\subseteq 2^C$. For each scene $s\in S$, $b_s$ and $e_s$ 
are the points in time when $s$ begins and ends, respectively. 
The time intervals $[b_{s_1},e_{s_1}]$ and $[b_{s_2},e_{s_2}]$ of two distinct scenes 
$s_1$ and $s_2$ may have a non-empty intersection, but if they do, 
we require $s_1\cap s_2=\emptyset$. 

The storyline visualization problem requires depicting each 
character $c\in C$ by a curve in the Euclidean plane that is strictly 
monotone on the time axis that we arbitrarily fix to the horizontal $x$-axis. 
The curve begins at the $x$-coordinate $x_c^b=\min\{b_s\mid c\in s\}$ and 
ends at $x_c^e=\max\{e_s\mid c\in s\}$. 
We call the interval $[x_c^b,x_c^e]$ the lifespan of character $c$. 

\begin{figure}[t]
  \centering
  \includegraphics[scale=0.9]{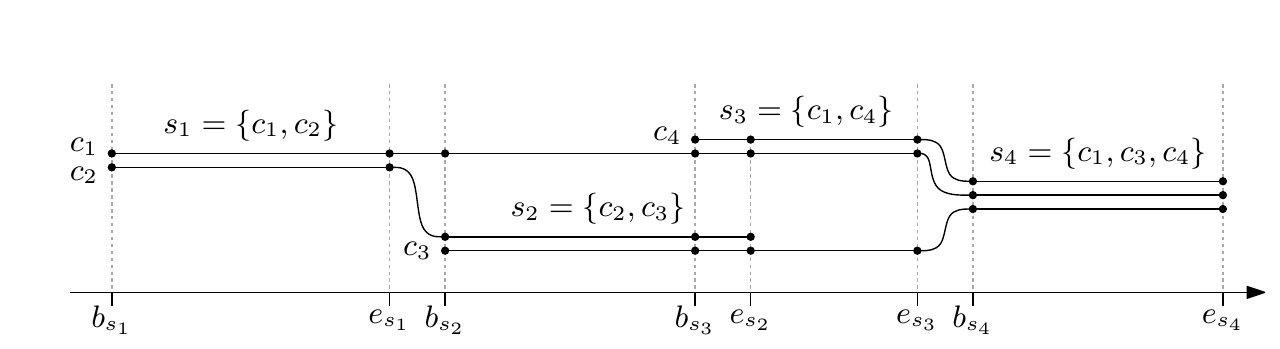}
  \caption{An example of a story with four scenes and four characters, where characters $c_3$ and $c_4$ enter late, character $c_2$ leaves early, and the time intervals $[b_{s_2},e_{s_2}]$ and $[b_{s_3},e_{s_3}]$ have a non-empty intersection.}
  \label{fig:story}
\end{figure}

The curves must be such that for every scene 
\hbox{$s=\{c_{\sigma_1},c_{\sigma_2},\ldots,c_{\sigma_k}\}\in S$} the $k$ corresponding 
curves in the interval $[b_s,e_s]$ are horizontal parallel lines 
that are equally spaced with vertical distance $1$. Furthermore, 
the curves of all $c\not\in s$ are restricted to $y$-coordinates that have 
an absolute difference of at least $2$ to the $y$-coordinates of the 
curves $c_{\sigma_i}\in s$ in the interval $[b_s,e_s]$, and to all curves for 
characters that are not members of any scene that intersects with $[b_s,e_s]$. 
An example is given in Fig.~\ref{fig:story}.

Given a story $(C,S,\{[b_s,e_s]\mid s\in S\})$, we construct an MLCM-TC instance $G=(V,E,\mathcal{T})$ as follows:
\begin{enumerate}
  \item Sort the points in time $\{b_s\mid s\in S\}\cup\{e_s\mid
    s\in S\}$ in non-decreasing order, 
  and let $\langle t_1,t_2,\ldots,t_p\rangle$ be the sorted sequence.
  \item Associate a layer $V_r$ with each $t_r$ ($r\in\{1,2,\ldots,p\}$), 
  create a node $v_{c,r}$ for each character $c$ for which $t_r$ is within its lifespan, 
  i.e., for which $t_r\in[x_c^b,x_c^e]$, and let $V_r=\{v_{c,r}\mid t_r\in[x_c^b,x_c^e]\}$.
  \item Let $V=\bigcup_{r=1}^p V_r$.
  \item Let $E=\big\{\{v_{c,r},v_{c,r+1}\}\text{ for all }c\in C\text{ such that }t_r,t_{r+1}\in[x_c^b,x_c^e]\big\}$.
  \item For each layer $V_r$ create a tree $T_r$ as follows:
  \begin{enumerate}[label=\roman*.]
    \item For each scene $s=\{c_{\sigma_1},c_{\sigma_2},\ldots,c_{\sigma_k}\}$ 
    such that $t_r\in [b_s,e_s]$ create an internal tree node $v_{s,r}$ 
    and tree edges $\{v_{s,r},v_{c_{\sigma_i},r}\}$ for all $i\in\{1,2,\ldots,k\}$.
    \item Unless the above results in a rooted tree with all nodes in $V_r$ as leaves, 
    create a tree root $\rho_r$ and tree edges connecting $\rho_r$ to all 
    previously created internal tree nodes of $T_r$ and to all character nodes in $V_r$ 
    that are not joined to a previously added internal tree node.
  \end{enumerate}
\end{enumerate}

\begin{figure}[t]
  \centering
  \includegraphics[scale=0.89]{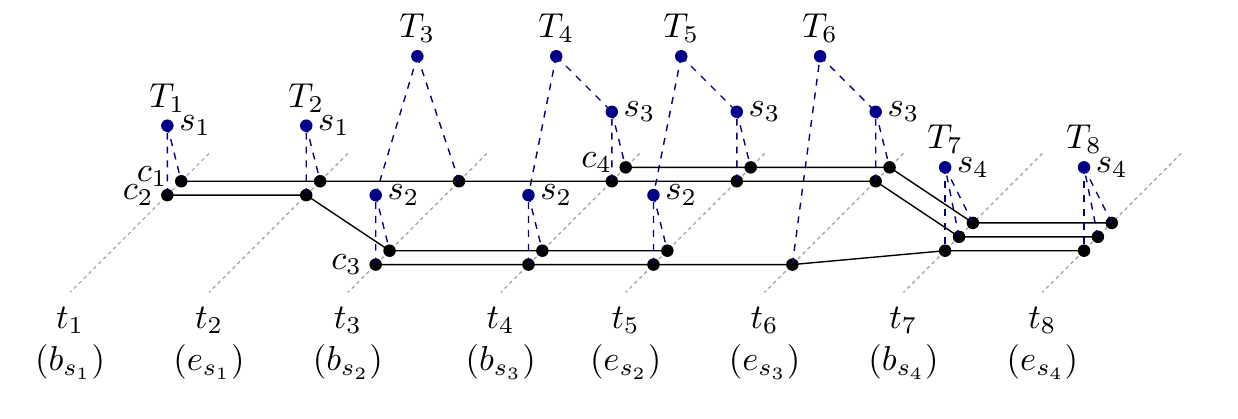}
  \caption{The MLCM-TC instance of the story of Fig.~\ref{fig:story}.}
  \label{fig:construction}
\end{figure}

\noindent In Fig.~\ref{fig:construction}, we demonstrate the construction for our example instance from
Fig.~\ref{fig:story}. 

Notice that the trees in $\mathcal{T}$ are all of height up to $2$, 
which means that storyline visualization instances yield a special subclass of MLCM-TC instances. 
By construction, an optimal solution of this MLCM-TC 
instance induces a storyline visualization with the minimum number of crossings, and, 
conversely, any instance of this special MLCM-TC subclass with trees of height up to $2$ 
is the result of the given transformation for some story. Thus, both problems are equivalent.
As MLCM can be reduced to this special subclass, NP-hardness is maintained.

%% file: ilpformulation.tex
\section{Integer Linear Programming Formulation}\label{sec:Model}
We present an integer linear programming (ILP) 
formulation of MLCM-TC. ILP
formulations have already been introduced for the general MLCM
problem~\cite{hk04,jlmo97} as well as for MLCM-TC, when restricted to the
special case of two layers only~\cite{bbl10}. Both models use
quadratic ordering formulations. In this section, we will extend these
formulations to an ILP model for MLCM-TC. 

To this end, let $G=(V,E,\mathcal{T})$ be an instance of MLCM-TC, as
described in Sect.~\ref{sec:SVtoMLCM}. For every layer
$r\in\{1,2,\ldots,p\}$, let \hbox{$V_r^{(2)}=\{(i,j)\in
V_r\times V_r:i<j\}$} be the set of all the ordered pairs of nodes on
the considered layer with the first index smaller than the second. 
As the total number of edge crossings is the sum of all crossings in
adjacent layers $r$ and $r+1$, summed up for all
\hbox{$r\in\{1,2,\ldots,p-1\}$,} let us consider the 
problem for a pair of adjacent layers $r$ and $r+1$, with \hbox{$r\in\{1,2,\ldots,p-1\}$.}

\begin{figure}[t]
  \centering
  \includegraphics{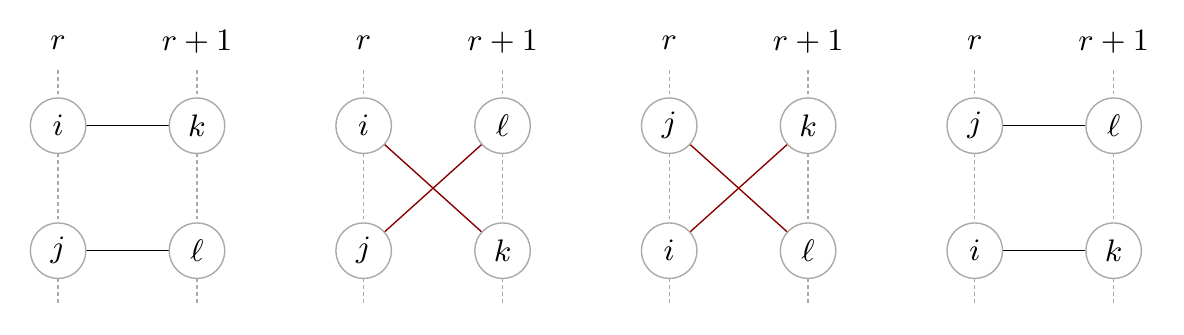}
  \caption{An edge pair crosses in two of four cases.}
  \label{fig:fourcases}
\end{figure}

A permutation of the nodes in $V_r$ is characterized by variables $x_{ij}^r\in\{0,1\}$ associated with the pairs $(i,j)\in V_r^{(2)}$ as follows:
\[x_{ij}^r=1\quad\text{if and only if $i$ is placed above $j$ on layer $r$}.\]
Then a pair of edges $(i,k),(j,\ell)\in E_r$ crosses if and only if
\[\text{$i$ is placed above $j$ on layer $r$ and $\ell$ is placed above $k$ on layer $r+1$}\]
or
\[\text{$j$ is placed above $i$ on layer $r$ and $k$ is placed above $\ell$ on layer $r+1$},\]
see Fig.~\ref{fig:fourcases}.

Therefore, if $\{x_{ij}^r\mid(i,j)\in V_r^{(2)}\}$ and $\{x_{k\ell}^{r+1}\mid(k,\ell)\in V_{r+1}^{(2)}\}$
describe node permutations on layers $r$ and $r+1$, respectively, we have
\[c_{ijk\ell}:=x_{ij}^r(1-x_{k\ell}^{r+1})+(1-x_{ij}^r)x_{k\ell}^{r+1}\in\{0,1\}\]
and $c_{ijk\ell}=1$ if and only if the edges $(i,k)$ and $(j,\ell)$ cross.

It is well known (see, e.g.,~\cite{gjr85}) that $\{x_{ij}^r\in\{0,1\}\mid(i,j)\in V_r^{(2)}\}$ 
characterizes a node permutation on $V_r$ if and only if the \emph{transitivity conditions}
\[0\le x_{hi}^r+x_{ij}^r-x_{hj}^r\le1\qquad(h<i<j)\]
are satisfied for all $r\in\{1,2,\ldots,p\}$.

It remains to model the tree conditions implied by the
elements of $\mathcal{T}$.
Given a layer $r\in\{1,2,\ldots,p\}$ and two nodes $i$
and $j$ in $V_r$, we denote by $P(i,j)$ the lowest common ancestor of
$i$ and $j$ in $T_r$. Let $V_r^{(3)}=\{(h,i,j)\in V_r\times V_r\times
V_r:h<i<j\}$. For every $r\in\{1,2,\ldots,p\}$ and every triple
$(h,i,j)\in V_r^{(3)}$, we impose the \emph{tree constraints}
\[\begin{array}{ll}
 x_{hj}^r=x_{ij}^r&\qquad\text{if }P(h,i)\neq P(P(h,i),j), \\
 x_{hi}^r=x_{hj}^r&\qquad\text{if }P(i,j)\neq P(h,P(i,j)).
 \end{array}\]
The first equation forbids the placement of $j$
between $h$ and $i$ in case $j$ does not belong to the smallest subtree
containing $h$ and $i$. Similarly, the second equation forbids the placement of $h$ between $i$ and
$j$ in case $h$ is not contained in the smallest subtree of $i$ and $j$. 

Putting it all together, we obtain the following model for MLCM-TC based on a combination of~\cite{jlmo97} for MLCM and~\cite{bbl10} for the special case of MLCM-TC for two layers:
\[\text{minimize}\qquad\sum\limits_{r=1}^{p-1}\;\sum\limits_{\stackrel{\scriptstyle{(i,j)\in V_r^{(2)},\;(k,\ell)\in V_{r+1}^{(2)}}}{(i,k),(j,\ell)\in E_r}}x_{ij}^r(1-x_{k\ell}^{r+1})+(1-x_{ij}^r)x_{k\ell}^{r+1}\]
subject to
\[\begin{array}{c@{\quad}ll}
0\;\le\;x_{hi}^r+x_{ij}^r-x_{hj}^r\le1&\text{for all }r\in\{1,2,\ldots,p\}&\text{ and }(h,i,j)\in V_r^{(3)}\\
x_{hj}^r=x_{ij}^r&\text{for all }r\in\{1,2,\ldots,p\}&\text{ and }(h,i,j)\in V_r^{(3)}\\
&&\text{ if }P(h,i)\neq P(P(h,i),j)\\
x_{hi}^r=x_{hj}^r&\text{for all }r\in\{1,2,\ldots,p\}&\text{ and
}(h,i,j)\in V_r^{(3)}\\
&&\text{ if }P(i,j)\neq P(h,P(i,j))\\
x_{ij}^r\in\{0,1\}&\text{for all }r\in\{1,2,\ldots,p\}&\text{ and }(i,j)\in V_r^{(2)}.
\end{array}\]

This is a quadratic 0-1-programming problem with linear constraints, namely, the transitivity conditions and the tree conditions. (Without the tree conditions, the problem is also called a \emph{quadratic linear ordering problem}.)

When we temporarily ignore the transitivity conditions and the tree conditions, the remaining problem is known as \emph{quadratic 0-1-optimization} of the form
\[\begin{array}{rl}
\text{minimize}&z^TQz+q^Tz\\
\text{s.t.}&z\in\{0,1\}^N
\end{array}\]
for an upper triangular matrix $Q\in\mathbb{Z}^{N\times N}$ and a vector $q\in\mathbb{Z}^N$. A well known construction of Hammer~\cite{ham65}, see also~\cite{bjr89,d89,ljrr04}, results in an equivalent formulation as a maximum cut problem on a graph $G_\textit{mc}=(V_ \textit{mc},E_ \textit{mc})$ with $N+1$ nodes, all but one are identified with the $z_i$, $i\in\{1,2,\ldots,N\}$. Let us call the additional node $z_0$, so $V_ \textit{mc}=\{z_0,z_1,\ldots,z_N\}$. The undirected edges $(z_i,z_j)$, $1\le i<j\le N$, correspond to the nonzero entries of the matrix $Q$, and there are additional $N$ edges $(z_0,z_i)$ for $1\le i\le N$, giving the edge set $E_ \textit{mc}$. The edge weights $w_e=w_{ij}$, $0\le i<j\le N$, are easily computed from $Q$ and $q$. For $W\subseteq V_ \textit{mc}$ the edge set $\delta(W)=\{(i,j)\in E_ \textit{mc}\mid i\in W,\,j\in V_ \textit{mc}\setminus W\}$ is called a \emph{cut} in $G_\textit{mc}$. Then the resulting \emph{maximum cut problem} has the form
\[\max\{w(\delta(W))\mid W\subseteq V_ \textit{mc}\}.\]

By introducing variables $y_e\in\{0,1\}$ for each $e\in E_\textit{mc}$, the maximum cut problem can be formulated as
\[\text{maximize}\qquad\sum\limits_{e\in E_\textit{mc}}w_ey_e\]
subject to
\[\begin{array}{c@{\quad}l}
\sum\limits_{e\in F}y_e-\sum\limits_{e\in C\setminus F}y_e\le|F|-1&\text{for all cycles }C\subseteq E_\textit{mc}\text{ and all }F\subseteq C,\,|F|\text{ odd}\\
y_e\in\{0,1\}&\text{for all }e\in E_\textit{mc}\,,
\end{array}\]
see~\cite{bm86}. The constraints are called \emph{odd cycle constraints}.

Applying this transformation is the key to our algorithm: 
The edges $e\in E_\textit{mc}$ not incident to $z_0$ correspond to edge pairs $(i,k),(j,\ell)\in E_r$, \hbox{$r\in\{1,2,\ldots,p-1\}$.}
The edges $e\in E_\textit{mc}$ that are incident to $z_0$ correspond to our variables $x_{ij}^r$ for \hbox{$r\in\{1,2,\ldots,p\}$, $i<j$. }
In view of the latter property, we can formulate \hbox{MLCM-TC} as a maximum
cut problem with the additional transitivity and tree
constraints, and we can solve it using a branch and cut approach for the maximum cut problem like in~\cite{bjr89} that additionally enforces these extra constraints. 

%% file: implementation.tex
\section{Implementation}\label{sec:Impl}

The implementation used to determine the minimum number of crossings in 
a storyline visualization consists of two main phases, a preprocessing phase and 
a branch and cut phase. During the preprocessing, we first reduce the number 
of layers of the problem (if possible), by identifying two consecutive layers $r$ and $r+1$
in case the corresponding trees $T_r$ and $T_{r+1}$ are identical and 
every node in $V_r$ and $V_{r+1}$ is an end of one edge of $E_r$ 
(e.g., layers $4$ and $5$ of Fig. \ref{fig:construction} can be identified). 
Then, a variant of the barycenter heuristic proposed by Sugiyama et al. \cite{stt81}, 
in which the presence of the trees on layers is taken into account, 
is executed 
in order to obtain an initial feasible solution that defines the indexing within the layers:
In this heuristic, the nodes of the trees are sorted according to their barycenters. 
The barycenter of a given leaf $t$ is computed by assigning to each edge, 
that has $t$ as end, the relative position of the other end as weight. 
The barycenter of each internal node $\tau$ is the mean of the 
barycenters of all the leaves of the subtree rooted at $\tau$. 

During the creation of the maximum cut graph induced by the heuristic solution, 
we exploit the fact that the tree constraints 
force many variables to assume the same value, so that we can identify them. 
Moreover, this procedure reduces also the number of constraints consistently 
after all variables have been replaced by their representatives: 
On the one hand, the tree constraints are not needed in the 
formulation anymore;  on the other hand, some transitivity constraints become 
deactivated or redundant. It is important to point out that, during this first phase, 
the problem is initialized without constraints and they are added according to need during 
the subsequent branch and cut phase. 

The branch and cut phase is realized in C++ using ABACUS~\cite{jt00} and CPLEX~\cite{cplex}.
The initial relaxation consists just of the objective function together with lower bounds~0  
and upper bounds~1 for the variables.
Odd cycle constraints and transitivity constraints are generated via separation, the 
former with the same strategy as described in~\cite{bjr89}, the latter by complete enumeration.

%% file: computationalresults.tex
\section{Computational Results}\label{sec:CompRes}

Our test-bed consists of:
\begin{itemize}
  \item[--] three movie instances \cite{movie}, namely ``Inception'', the original trilogy of ``Star Wars'' and ``The Matrix'';
  \item[--] three book instances from the Stanford GraphBase database \cite{sgb}, namely ``Adventures of Huckleberry Finn'', ``Anna Karenina'' and ``Les Mis\'erables''.
\end{itemize}

These instances have been converted to MLCM-TC by using the procedure described in Sect. \ref{sec:SVtoMLCM}. 
In the conversion of the book instances, a slight change is required: Since these instances do not report time intervals, 
but just the list of the characters involved in each scene of each chapter, a layer has been created for each of these scenes, 
instead of for each beginning and ending time point.

The three movie instances have been generated using the raw data set from \cite{movie} in order
to compare them with results in the literature. We obtained
``Inception'', ``Star Wars'' and ``The Matrix'' following the
principles described in Sect.~\ref{sec:SVtoMLCM}. 
However, after having solved them, we realized that the number of
crossings given by our algorithm for ``Inception'' was $35$, while it was $24$
in \cite{thm15} and $23$ in \cite{storyflow13}. After a careful study of the layouts provided
in~\cite{storyflow13,thm15,tm12}, we noticed that the storylines of
``Inception'' and ``The Matrix'' in~\cite{storyflow13,thm15,tm12} differ from the raw data set
provided by \cite{movie}, and therefore are not comparable with our instances. 

In order to make a
comparison possible, ``Inception'' required three major modifications.
This modified instance is called ``Inception-sf'' and is generated by incorporating the following changes
that are based on a careful study of the layouts provided in \cite{storyflow13,thm15,tm12}.
The storyline for the character ``Mal'' is allowed to take shortcuts, i.e., in long periods of absence
it is drawn as a thin curve that may cross other storylines without accounting for these crossings 
(see Fig. 12 in~\cite{storyflow13}). Moreover, the grouping at the end of the movie does not
correspond to the last scene in the data set. To keep our layout comparable, we enforced in our new instance
the same grouping at the end. The third discrepancy is the number of characters. In the data from~\cite{movie}
there are ten characters listed in the corresponding file, 
whereas the layouts from the literature~\cite{storyflow13,thm15,tm12} contain only eight storylines, in which 
``Arch'' and ``Asian'' are missing. A major modification was also
necessary in ``The Matrix'', where the
storylines for the characters ``Brown'', ``Smith'' and ``Jones'' are
allowed to take shortcuts as well. We call it ``The Matrix-sf''.

Since the instances ``Anna Karenina'' and ``Les Mis\'erables'' are very big, 
we have split them into chapters and sequences of chapters. 
The resulting test-bed is made of eight chapters, seven pairs of chapters, 
six triples of chapters and five quadruples of chapters from ``Anna Karenina'', 
and five chapters, four pairs of chapters and three triples of chapters 
from ``Les Mis\'erables'', plus the entire ``Adventures of Huckleberry Finn'', 
``Inception-sf'', ``Inception'', ``Star Wars'', ``The Matrix-sf'', and ``The Matrix''.

To the best of our knowledge, this is the first time in which ILP techniques are applied to storyline visualizations. 
Thus comparisons of computational results are not possible. 
Runs were performed on one node of the HPC Cluster of the Computer Science Department of the University of Cologne. 
The node used consists of two Intel E5-2690v2 CPUs with ten cores each and 128GB RAM. 

While the book instances generated from the Stanford GraphBase database are introduced here for the first time, 
the literature provides crossing counts for the three movie instances (``Inception'', ``Star Wars'', and ``The Matrix'').
Table~\ref{tb:movie_comparision} shows a comparison
of the minimum number of crossings (OPT) from our approach with the numbers of crossings obtained by 
 the streaming-oriented approach from Tanahashi et al.~\cite{thm15} (THM),
the Storyflow approach from Liu et al.~\cite{storyflow13} (LIU), and 
the evolutionary algorithm from Tanahashi and Ma~\cite{tm12}
(TM). Crossing counts for THM, LIU and TM are taken from Table 3 in \cite{thm15}.
We can confirm that the best solution reported by Liu et al.~\cite{storyflow13} for the movie ``Inception'' is optimal. 
For ``Star Wars'' the approach from Tanahashi et al.~\cite{thm15} comes very close to the optimal solution,
even though the instance is the biggest and has the highest crossing count.
One may conclude that the heuristics in~\cite{storyflow13,thm15} deliver solutions with a good crossing count, 
especially when considering the fact that they do not optimize the crossing count alone.

\begin{table}[h]
 \caption{Comparison of the solution of the movies.}
 \label{tb:movie_comparision}
 \centering
 \setlength{\tabcolsep}{2.00mm}
 \begin{tabular}{|l|cccc|}
 \hline
 & \multicolumn{1}{|c}{OPT} & \multicolumn{1}{c}{THM~\cite{thm15}} & \multicolumn{1}{c}{LIU~\cite{storyflow13}} & \multicolumn{1}{c|}{TM~\cite{tm12}} \\
 \hline
 Inception-sf & $\bf 23$ & $24$ & $\bf 23$ & $99$ \\
 Star Wars & $\bf 39$ & $\bf 41$ & $48$ & $51$ \\
 The Matrix-sf & $\bf 10$ & $22$ & $\bf 14$ & $43$ \\
 \hline
 \end{tabular}
 \label{tab:Movies}
\end{table}

\begin{table}[p]
  \caption{Information about the solution of the considered instances.}
  \centering
  \setlength{\tabcolsep}{0.78mm}
  \begin{tabular}{|l|rrr|rrrrrrr|}
  \hline
  & \multicolumn{1}{|c}{$p$} & \multicolumn{1}{c}{$|V|$} & \multicolumn{1}{c|}{$|E|$} & \multicolumn{1}{|c}{cr} & \multicolumn{1}{c}{$n_\textit{var}$} & \multicolumn{1}{c}{$n_\textit{oddc}$} & \multicolumn{1}{c}{$n_\textit{trans}$} & \multicolumn{1}{c}{$n_\textit{sub}$} & \multicolumn{1}{c}{$n_\textit{LPs}$} & Time \\
  \hline
  anna1 & $58$ & $409$ & $368$ & $\textbf{20}$ & $1\,944$ & $2\,684$ & $60$ & $31$ & $344$ & $13.03$ \\
  anna2 & $58$ & $525$ & $489$ & $\textbf{12}$ & $3\,689$ & $2\,665$ & $1$ & $1$ & $126$ & $0.88$ \\
  anna3 & $48$ & $265$ & $219$ & $\textbf{0}$ & $951$ & $0$ & $0$ & $1$ & $1$ & $0.01$ \\
  anna4 & $49$ & $364$ & $334$ & $\textbf{20}$ & $2\,116$ & $2\,231$ & $48$ & $13$ & $159$ & $4.86$ \\
  anna5 & $71$ & $615$ & $565$ & $\textbf{17}$ & $3\,821$ & $3\,182$ & $60$ & $3$ & $197$ & $2.60$ \\
  anna6 & $56$ & $522$ & $495$ & $\textbf{31}$ & $3\,586$ & $4\,368$ & $49$ & $3$ & $150$ & $3.89$ \\
  anna7 & $62$ & $467$ & $420$ & $\textbf{9}$ & $2\,525$ & $2\,278$ & $82$ & $17$ & $191$ & $7.88$ \\
  anna8 & $28$ & $192$ & $175$ & $\textbf{6}$ & $1\,036$ & $850$ & $1$ & $1$ & $45$ & $0.15$ \\
  \hline
  anna1-2 & $117$ & $1\,454$ & $1\,397$ & $\textbf{57}$ & $16\,433$ & $18\,284$ & $89$ & $5$ & $545$ & $196.24$ \\
  anna2-3 & $108$ & $1\,461$ & $1\,394$ & $\textbf{28}$ & $18\,763$ & $16\,849$ & $29$ & $3$ & $469$ & $48.96$ \\
  anna3-4 & $100$ & $1\,015$ & $951$ & $\textbf{34}$ & $8\,473$ & $8\,516$ & $45$ & $3$ & $328$ & $12.66$ \\
  anna4-5 & $126$ & $1\,808$ & $1\,748$ & $\textbf{78}$ & $23\,742$ & $26\,129$ & $181$ & $3$ & $814$ & $306.32$ \\
  anna5-6 & $129$ & $1\,760$ & $1\,697$ & $\textbf{76}$ & $19\,967$ & $23\,155$ & $252$ & $3$ & $656$ & $281.26$ \\
  anna6-7 & $120$ & $1\,445$ & $1\,385$ & $\textbf{79}$ & $14\,464$ & $32\,396$ & $671$ & $5$ & $3\,008$ & $1\,387.57$ \\
  anna7-8 & $90$ & $905$ & $850$ & $\textbf{32}$ & $7\,248$ & $8\,711$ & $265$ & $3$ & $365$ & $19.16$ \\
  \hline
  anna1-3 & $166$ & $2\,948$ & $2\,865$ & $[100,199]$ & $52\,072$ & $61\,743$ & $631$ & $1$ & $1\,155$ & t.l. \\
  anna2-4 & $158$ & $2\,637$ & $2\,557$ & $\textbf{78}$ & $40\,789$ & $46\,600$ & $351$ & $3$ & $2\,042$ & $1\,284.03$ \\
  anna3-5 & $174$ & $3\,100$ & $3\,012$ & $[115,224]$ & $51\,814$ & $60\,646$ & $366$ & $7$ & $1\,391$ & t.l. \\
  anna4-6 & $178$ & $3\,115$ & $3\,044$ & $[124,298]$ & $50\,106$ & $207\,148$ & $232$ & $3$ & $1\,697$ & t.l. \\
  anna5-7 & $191$ & $3\,742$ & $3\,656$ & $[144,361]$ & $69\,156$ & $77\,742$ & $653$ & $1$ & $1\,216$ & t.l. \\
  anna6-8 & $146$ & $2\,205$ & $2\,140$ & $[117,200]$ & $28\,767$ & $45\,396$ & $864$ & $3$ & $2\,052$ & t.l. \\
  \hline
  anna1-4 & $216$ & $4\,627$ & $4\,534$ & $[115,339]$ & $98\,525$ & $100\,149$ & $251$ & $1$ & $1\,252$ & t.l. \\
  anna2-5 & $232$ & $5\,366$ & $5\,266$ & $[102,350]$ & $116\,249$ & $111\,255$ & $261$ & $1$ & $1\,001$ & t.l. \\
  anna3-6 & $226$ & $5\,262$ & $5\,168$ & $[122,424]$ & $119\,573$ & $121\,148$ & $180$ & $1$ & $1\,345$ & t.l. \\
  anna4-7 & $240$ & $5\,467$ & $5\,375$ & $[117,504]$ & $119\,974$ & $123\,020$ & $238$ & $1$ & $1\,166$ & t.l. \\
  anna5-8 & $217$ & $4\,624$ & $4\,534$ & $[123,470]$ & $93\,832$ & $97\,792$ & $377$ & $1$ & $1\,088$ & t.l. \\
  \hline
  huck & $107$ & $1\,059$ & $985$ & $\textbf{42}$ & $7\,942$ & $11\,024$ & $357$ & $29$ & $1\,098$ & $111.31$ \\
  \hline
  jean1 & $95$ & $502$ & $462$ & $\textbf{10}$ & $1\,777$ & $1\,265$ & $49$ & $3$ & $167$ & $0.90$ \\
  jean2 & $59$ & $226$ & $212$ & $\textbf{6}$ & $461$ & $385$ & $0$ & $1$ & $44$ & $0.08$ \\
  jean3 & $99$ & $873$ & $838$ & $\textbf{13}$ & $6\,559$ & $3\,407$ & $801$ & $7$ & $360$ & $6.31$ \\
  jean4 & $76$ & $909$ & $876$ & $\textbf{42}$ & $9\,219$ & $10\,116$ & $177$ & $3$ & $335$ & $22.22$ \\
  jean5 & $73$ & $491$ & $471$ & $\textbf{17}$ & $2\,608$ & $2\,412$ & $4$ & $3$ & $138$ & $1.52$ \\
  \hline
  jean1-2 & $154$ & $1\,102$ & $1\,055$ & $\textbf{20}$ & $5\,823$ & $4\,172$ & $111$ & $3$ & $226$ & $3.93$ \\
  jean2-3 & $159$ & $1\,808$ & $1\,767$ & $\textbf{33}$ & $18\,882$ & $14\,128$ & $1\,512$ & $3$ & $732$ & $48.90$ \\
  jean3-4 & $176$ & $3\,249$ & $3\,208$ & $[115,232]$ & $57\,746$ & $66\,222$ & $482$ & $1$ & $1\,698$ & t.l. \\
  jean4-5 & $149$ & $1\,943$ & $1\,907$ & $\textbf{96}$ & $24\,584$ & $32\,573$ & $619$ & $3$ & $1\,037$ & $1012.44$ \\
  \hline
  jean1-3 & $254$ & $2\,853$ & $2\,780$ & $\textbf{53}$ & $27\,720$ & $20\,886$ & $1\,991$ & $3$ & $1\,177$ & $143.34$ \\
  jean2-4 & $235$ & $4\,182$ & $4\,135$ & $[130,302]$ & $75\,150$ & $81\,236$ & $429$ & $1$ & $1\,928$ & t.l. \\
  jean3-5 & $248$ & $4\,429$ & $4\,386$ & $[101,372]$ & $79\,208$ & $83\,279$ & $503$ & $1$ & $1\,529$ & t.l. \\
  \hline
  Inception-sf & $137$ & $798$ & $787$ & $\textbf{23}$ & $1\,401$ & $1\,756$ & $7$ & $3$ & $108$ & $0.89$ \\
  Inception & $139$ & $925$ & $915$ & $\textbf{35}$ & $1\,784$ & $2\,376$ & $6$ & $3$ & $130$ & $2.02$ \\
  Star Wars & $100$ & $940$ & $926$ & $\textbf{39}$ & $2\,132$ & $2\,441$ & $8$ & $1$ & $168$ & $0.99$ \\
  The Matrix-sf & $82$ & $678$ & $660$ & $\textbf{10}$ & $1\,343$ & $1\,219$ & $18$ & $3$ & $125$ & $0.72$ \\
  The Matrix & $82$ & $683$ & $669$ & $\textbf{12}$ & $1\,388$ & $1\,328$ & $45$ & $3$ & $98$ & $0.77$ \\
  \hline
  \end{tabular}
  \label{tab:CompRes}
\end{table}

\begin{figure}[ht]
  \centering
  \begin{minipage}[b]{\textwidth}
        \centering
        \subfloat[The movie ``Inception'' with $35$ crossings.]{
        \includegraphics[width=0.95\textwidth]{fig/sv_inception.pdf}
        }
  \end{minipage}
  \myspace{0.4cm}
  \begin{minipage}[b]{\textwidth}
        \centering
        \subfloat[The movie ``Inception-sf'' with $23$ crossings.]{
        \includegraphics[width=0.95\textwidth]{fig/sv_inception-sf.pdf}
        }
  \end{minipage}
  \myspace{0.4cm}
  \begin{minipage}[b]{\textwidth}
        \centering
        \subfloat[The original trilogy of the movie ``Star Wars'' with $39$ crossings.]{
        \includegraphics[width=0.95\textwidth]{fig/sv_starwars.pdf}
        }
  \end{minipage}
  \myspace{0.4cm}
  \begin{minipage}[b]{\textwidth}
        \centering
        \subfloat[The movie ``The Matrix'' with $12$ crossings.]{
        \includegraphics[width=0.95\textwidth]{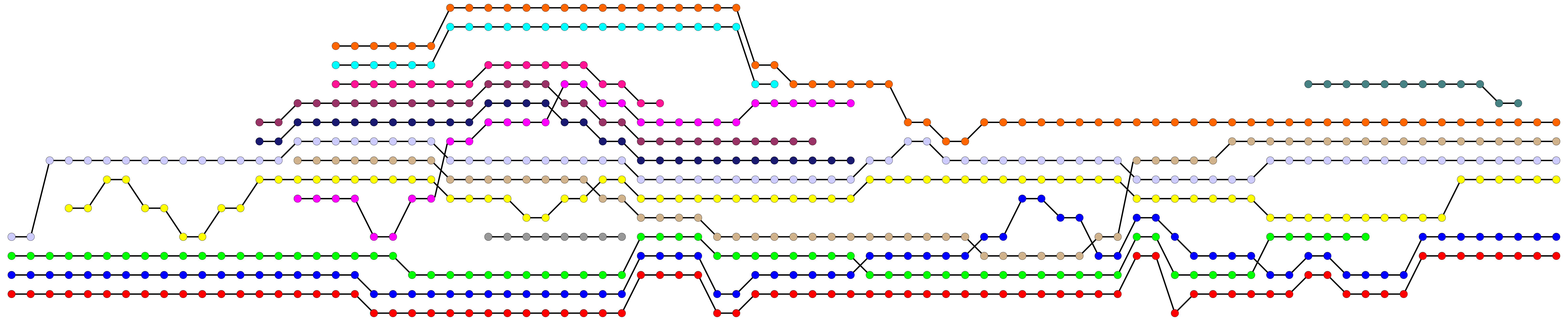}
        }
  \end{minipage}
  \caption{Storyline visualizations with minimum number of crossings of the three movies from~\cite{movie}.}
  \label{fig:svmovies}
\end{figure}

In Table \ref{tab:CompRes}, we report the information about the solution of the considered instances: 
The number of layers ($p$), of nodes ($|V|$), of edges ($|E|$), 
the minimum number of crossings (cr) in boldface or a pair $[$lower
bound, best known number of crossings$]$,
the number of variables ($n_{var}$), of odd cycle constraints added during 
the separation ($n_\textit{oddc}$), of transitivity constraints added during the separation ($n_\textit{trans}$), 
of subproblems in the branch and cut tree ($n_\textit{sub}$), of linear programming relaxations solved ($n_\textit{LPs}$), 
and the runtime expressed in seconds (Time) where ``t.l.'' means that the run was aborted due to the time limit of one hour, in which cases the cr column contains an interval.
While $29$ of the $42$ instances have been solved to optimality, for the remaining $13$ instances the best lower bound for the number of crossings differs from the best solution found at timeout termination.
 
When we analyze the behaviour of our algorithm, we have to distinguish between 
movie and book instances: Since the original instances from \cite{movie} 
allow more than one scene per layer, the trees on the layers of the movie 
instances restrict consistently the possible permutations of the corresponding nodes 
and consequently reduce the number of variables. On the other hand, 
this is not the case for the book instances, where only one scene per layer occurs. 
We can observe that MLCM-TC for movies tends to be much 
easier in comparison to a book instance with similar numbers of layers, 
nodes, and edges. 

The difficulty of a book instance is mainly influenced by the combination 
of two parameters: the number of layers $p$ and the number of nodes $|V|$. 
If the number of nodes is fixed, the higher the number of layers is, 
the easier the solution is, since the distribution of the nodes on more layers 
reduces the number of variables of the problem. On the other hand, 
if the number of layers is fixed, the difficulty increases with the number of nodes. 

The hardest instance we have been able to solve to optimality is
``anna2-4'', where $2\,637$ nodes are distributed on 
only $158$ layers which results in $40\,789$ variables. 
The biggest solved instance 
in terms of number of layers is ``jean1-3'' with $254$ layers but only $2\,853$ nodes, 
which results in $27\,720$ variables.

\begin{figure}[ht]
  \centering
  \begin{minipage}[b]{\textwidth}
        \centering
        \subfloat[The third chapter of the book ``Anna Karenina'' with $0$ crossings.]{
        \includegraphics[width=0.95\textwidth]{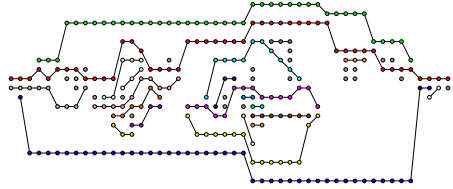}
        }
  \end{minipage}
  \myspace{0.4cm}
  \begin{minipage}[b]{\textwidth}
        \centering
        \subfloat[The first chapter of the book ``Les Mis\'erables'' with $10$ crossings.]{
        \includegraphics[width=0.95\textwidth]{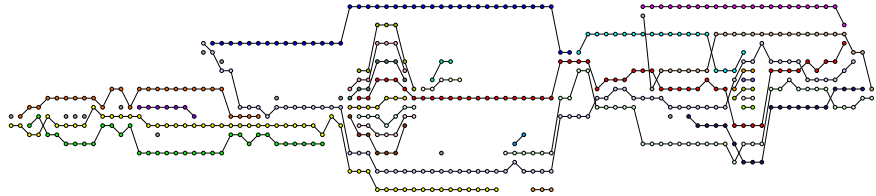}
        }
  \end{minipage}
  \caption{Storyline visualizations of two chapters from ``Anna Karenina'' and ``Les Mis\'erables''~\cite{sgb}.}
  \label{fig:svbooks}
\end{figure}

We present crossing minimal storyline visualizations of the three
movie instances in Fig.~\ref{fig:svmovies} and the two book instances in Fig.~\ref{fig:svbooks}.

%% file: conclusion.tex
\section{Conclusion}

In this work we have tackled the crossing minimization problem
in storyline visualization via an ILP formulation. Despite being an
NP-hard problem, computational results show that with our approach 
one can handle instances of medium size within a reasonable time frame.
However, our approach is of purely combinatorial nature, thus, extending
it to automatically generate storyline visualizations such that
other design criteria are taken into account is not straightforward.
%

%% file: acknowledments.tex
\section*{Acknowledgments}\label{sec:Acknowledgments}
The authors are grateful to K\"ate Zimmer who made her MLCM
code, developed in the context of her Master's thesis~\cite{z13},
available to us. Her code served as the basis for our experimental MLCM-TC implementation.
Our work is supported by the EU grant FP7-PEOPLE-2012-ITN -
Marie-Curie Action ``Initial Training Networks'' no.~316647
``Mixed-Integer Nonlinear Optimization'' (MINO).
